\documentclass[prd,superscriptaddress,showpacs,nofootinbib,showkeys]{revtex4}

\usepackage{graphicx,amsmath,amssymb,amsxtra,appendix}

\begin{document}

\title{Charged Fermion Masses and Mixing from a $SU(3)$ Family Symmetry Model}

\author{Albino Hern\'andez-Galeana}

\email{albino@esfm.ipn.mx}

\affiliation{ Departamento de F\'{\i}sica,   Escuela Superior de
F\'{\i}sica y Matem\'aticas, I.P.N., \\
U. P. "Adolfo L\'opez Mateos". C. P. 07738, M\'exico, D.F., M\'exico. }



\begin{abstract}
Within the framework of a Beyond Standard Model (BSM) with a local $SU(3)$ family symmetry, we report an updated fit of parameters which account for the known spectrum of quarks and charged lepton masses
and the quark mixing in a $4\times 4$ non-unitary $V_{CKM}$. In this scenario, ordinary heavy fermions, top and bottom quarks and tau lepton, become massive at tree level from Dirac See-saw mechanisms
implemented by the introduction of a new set of $SU(2)_L$ weak singlet vector-like
fermions, $U,D,E,N$, with $N$ a sterile neutrino. The $N_{L,R}$ sterile neutrinos allow
the implementation of a $8\times 8$ general See-saw Majorana neutrino mass
matrix with four massless eigenvalues at tree level. Hence, light fermions, including
neutrinos, obtain masses from loop radiative corrections mediated
by the massive $SU(3)$ gauge bosons.  $SU(3)$ family symmetry is broken spontaneously in two stages, whose hierarchy of scales yield an approximate  $SU(2)$ global symmetry associated with the  $Z_1, Y_1^\pm$  gauge boson masses of the order of 2 TeV. A global fit of parameters to include neutrino masses and lepton mixing is in progress.
 \end{abstract}

\maketitle

\tableofcontents

\section{ Introduction }

\vspace{3mm}
The origen of the hierarchy of fermion masses and mixing is one of the most important open
problems in particle physics. Any attempt to account for this hierarchy introduce a mass generation mechanism which distinguish among the different Standard Model (SM) quarks and leptons.

\vspace{3mm}

After the discovery of the scalar Higgs boson on 2012,  LHC has not found a conclusive evidence of new physics. However, there are theoretical motivations to look for new particles in order to answer some open questions like; neutrino oscillations, dark matter, stability of the Higgs mass against radiative corrections,etc.

\vspace{3mm}
In this report, we address the problem of charged fermion masses and
quark mixing within the framework of an extension of the SM introduced by the
author in \cite{albinosu32004}. This BSM proposal include a vector gauged
$SU(3)$ family symmetry\footnote{See \cite{albinosu32004,albinosu3bled} and references therein for some other $SU(3)$ family symmetry model proposals.} commuting with the SM group and introduce a hierarchical mass generation mechanism in which the light fermions obtain masses
through loop radiative corrections, mediated by the massive
bosons associated to the $SU(3)$ family symmetry that is
spontaneously broken, while the masses of the top and bottom
quarks and that of the tau lepton are generated at tree level
from "Dirac See-saw"\cite{SU3MKhlopov} mechanisms through the introduction of a new set of $SU(2)_L$ weak singlets $U,D,E$ and $N$ vector-like fermions, which do not couple to the $W$ boson, such that
the mixing of $U$ and $D$ vector-like quarks with the SM quarks gives rise to and extended $4\times4$ non-unitary CKM quark mixing matrix \cite{vectorlikepapers}.

\section{Model with $SU(3)$ flavor symmetry}

\subsection{Fermion content}

Before "Electroweak Symmetry Breaking"(EWSB) all ordinary SM fermions
remain massless, and the global symmetry in this limit, including
R-handed neutrinos, is:

\begin{eqnarray}
SU(3)_{q_L}\otimes SU(3)_{u_R}\otimes SU(3)_{d_R}\otimes
SU(3)_{l_L}\otimes SU(3)_{\nu_R}\otimes SU(3)_{e_R} \nonumber\\ \nonumber\\
\supset SU(3)_{q_L+u_R+d_R+l_L+e_R+\nu_R} \equiv SU(3) \label{su3symmetry}
\end{eqnarray}

\vspace{3mm}
\noindent We define the gauge symmetry group

\begin{equation}
G\equiv SU(3) \otimes G_{SM} \label{gaugegroup}
\end{equation} 

\noindent  where $SU(3)$ is the gauged family symmetry among families, eq.(\ref{su3symmetry}) , and
$G_{SM}= SU(3)_C \otimes SU(2)_L \otimes U(1)_Y $ is the "Standard Model" gauge group, with $g_H$, $g_s$, $g$ and $g^\prime$ the corresponding coupling constants. The content of fermions assumes the ordinary quarks and leptons assigned under G as:

\vspace{3mm}
{\bf Ordinary Fermions:}  $q_{iL}^o=\begin{pmatrix} u_{iL}^o \\ d_{iL}^o \end{pmatrix} \;,\;
l_{iL}^o=\begin{pmatrix} \nu_{iL}^o \\ e_{iL}^o \end{pmatrix}  \; , \; Q = T_{3L} + \frac{1}{2} Y$

\begin{equation*}
\Psi_q^o = ( 3 , 3 , 2 , \frac{1}{3} )_L=\begin{pmatrix} q_{1L}^o \\ q_{2L}^o \\ q_{3L}^o \end{pmatrix}  \quad , \quad \Psi_l^o= ( 3 , 1 , 2 , -1 )_L=\begin{pmatrix} l_{1L}^o \\ l_{2L}^o \\ l_{3L}^o \end{pmatrix} \end{equation*}

\begin{equation*}
\Psi_u^o = ( 3 , 3, 1 , \frac{4}{3} )_R=\begin{pmatrix} u_R^o \\ c_R^o \\ t_R^o \end{pmatrix} \quad ,
\quad \Psi_d^o =(3, 3 , 1 , -\frac{2}{3} )_R=\begin{pmatrix} d_R^o \\ s_R^o \\ b_R^o \end{pmatrix} \end{equation*}

\begin{equation*}
\Psi_e^o = (3 , 1 , 1,-2)_R=\begin{pmatrix} e_R^o \\ \mu_R^o \\ \tau_R^o \end{pmatrix} \: ,
\end{equation*}

\noindent where the last entry corresponds to the hypercharge $Y$. The model also includes two types of extra $SU(2)_L$ weak singlet fermions:

\noindent {\bf Right Handed Neutrinos:} $ \Psi_{\nu_R}^o = ( 3 , 1 , 1 , 0 )_R= \begin{pmatrix} 
 \nu_{e_R} \\  \nu_{\mu_R} \\    
\nu_{\tau_R}  \end{pmatrix}  $ , 

\noindent and the vector-like fermions:

\vspace{4mm}
\noindent {\bf Sterile Neutrinos: } $\quad N_L^o, N_R^o = ( 1 , 1 , 1 , 0 )  $ , 

\vspace{5mm}

\noindent {\bf The Vector Like quarks:}
\begin{equation} 
U_L^o, U_R^o = ( 1 , 3 , 1 , \frac{4}{3} ) \quad , \quad
D_L^o, D_R^o = ( 1 , 3 , 1 ,- \frac{2}{3} )  \label{vectorquarks} 
\end{equation} 

\noindent and 
\vspace{3mm}

\noindent {\bf The Vector Like electron:} $\quad E_L^o, E_R^o = ( 1 , 1 , 1 , -2 ) $.\\

\vspace{4mm}
\noindent The transformation of these vector-like fermions allows the gauge invariant mass terms

\begin{equation}
M_U \:\bar{U}_L^o \:U_R^o \,+\, M_D \:\bar{D}_L^o \:D_R^o \,+\, M_E \:\bar{E}_L^o \:E_R^o + h.c. \;,
\end{equation}

and

\begin{equation}
m_D \,\bar{N}_L^o \,N_R^o \,+\, m_L \,\bar{N}_L^o\, (N_L^o)^c \,+\, m_R \,\bar{N}_R^o\, (N_R^o)^c \,+\,  h.c
\end{equation}

\noindent The above fermion content
make the model anomaly free. After the definition of the gauge
symmetry group and the assignment of the ordinary fermions in the
usual form under the standard model group and in the
fundamental $3$-representation under the $SU(3)$ family symmetry,
the introduction of the right-handed neutrinos is required to
cancel anomalies \cite{T.Yanagida1979}. The $SU(2)_L$ weak singlet
vector-like fermions have been introduced to give masses at tree
level only to the third family of known fermions via Dirac
See-saw mechanisms. These vector like fermions, together with the radiative corrections, 
play a crucial role to implement a hierarchical spectrum for ordinary quarks and charged lepton
masses.

\section{$SU(3)$ family symmetry breaking}

To implement a hierarchical spectrum for charged fermion masses,
and simultaneously to achieve the SSB of $SU(3)$, we introduce the
flavon scalar fields: $\eta_i,\;i=2,3$, 

\begin{equation*}
\eta_i=(3 , 1 , 1 , 0)=\begin{pmatrix} \eta_{i1}^o\\ \eta_{i2}^o\\ \eta_{i3}^o \end{pmatrix} \;, \quad i=2,3,
\end{equation*}

\noindent acquiring the "Vacuum ExpectationValues" (VEV's):

\begin{equation}
\langle \eta_2 \rangle^T = ( 0 , \Lambda_2 , 0) \quad , \quad
\langle \eta_3 \rangle^T = ( 0 , 0,  \Lambda_3)  \:. \label{veveta2eta3} \end{equation}

\noindent 
The corresponding $SU(3)$ gauge bosons
are defined in Eq.\eqref{SU3lagrangian} through their couplings to
fermions. Thus, the contribution to the horizontal gauge boson masses
from Eq.(\ref{veveta2eta3}) read

\begin{itemize}

\item $\eta_2:\quad \frac{g_{H_2}^2 \Lambda_2^2}{2} ( Y_1^+ Y_1^- + Y_3^+ Y_3^-) +  \frac{g_{H_2}^2 \Lambda_2^2}{4} ( Z_1^2 + \frac{Z_2^2}{3} - 2 Z_1 \frac{Z_2}{ \sqrt{3}} ) $

\item $\eta_3:\quad \frac{g_{H_3}^2 \Lambda_3^2}{2} ( Y_2^+ Y_2^- + Y_3^+ Y_3^-) + g_{H_3}^2 \Lambda_3^2 \frac{Z_2^2}{3} $
\end{itemize}

\noindent  {\it  These two scalars in the fundamental representation is the minimal set of scalars to break down completely  the $SU(3)$ family symmetry}. Therefore, neglecting tiny contributions from electroweak symmetry breaking, Eq.\eqref{ewyimixcont}, we obtain the gauge boson mass terms:

\begin{equation}
 M_2^2 \,Y_1^+ Y_1^- + M_3^2 \,Y_2^+ Y_2^- + ( M_2^2 + M_3^2) \,Y_3^+ Y_3^- 
+ \frac{1}{2} M_2^2  \,Z_1^2 +\frac{1}{2} \frac{M_2^2 + 4 M_3^2}{3} \,Z_2^2  
- \frac{1}{2}( M_2^2 ) \frac{2}{\sqrt{3}}  \,Z_1 \,Z_2
\end{equation}

\begin{equation} M_2^2= \frac{g_{H}^2 \Lambda_2^2}{2} \quad , \quad M_3^2=\frac{g_{H}^2 \Lambda_3^2}{2}  \quad , \quad y \equiv \frac{M_3}{M_2}= \frac{\Lambda_3}{\Lambda_2} \label{M23}
\end{equation}

\begin{table}[!]
\begin{center} \begin{tabular}{ c | c c }
   &  $Z_1$ & $Z_2$ \\
\hline    \\
$Z_1$ &   $ M_2^2$ &  $ - \frac{  M_2^2}{\sqrt{3}}$ \\
      &           &                            \\
$Z_2$  & $ - \frac{M_2^2}{\sqrt{3}}$  & $\quad \frac{M_2^2+4 M_3^2}{3}$
\end{tabular} \end{center}
\caption{$Z_1 - Z_2$ mixing mass matrix }
\end{table}

\noindent Diagonalization of the $Z_1-Z_2$ squared mass matrix yield the eigenvalues

\begin{eqnarray}
M_-^2=\frac{2}{3} \left( M_2^2 + M_3^2 - \sqrt{ (M_3^2 -  M_2^2)^2 + M_2^2  M_3^2 } \right)=M_2^2 \,y_-\label{Mm} \\
\nonumber\\
M_+^2=\frac{2}{3} \left( M_2^2 + M_3^2 +\sqrt{ (M_3^2 -  M_2^2)^2 + M_2^2  M_3^2 } \right)=M_2^2 \,y_+  \label{Mp}
\end{eqnarray}

\noindent and the gauge boson mass eigenvalues

\begin{equation}
M_2^2 \,Y_1^+ Y_1^- +  M_3^2\,Y_2^+ Y_2^- + ( M_2^2 + M_3^2) \,Y_3^+ Y_3^-
+ M_-^2 \,\frac{Z_-^2}{2} +  M_+^2 \,\frac{Z_+^2}{2}
\end{equation}

\noindent or

\begin{equation}
M_2^2 \,Y_1^+ Y_1^- +  M_2^2\,y^2\,Y_2^+ Y_2^- + M_2^2  ( 1 + y^2) \,Y_3^+ Y_3^-
+ M_2^2\,y_- \,\frac{Z_-^2}{2} +  M_2^2\,y_+ \,\frac{Z_+^2}{2}\, ,
\end{equation}

\noindent where

\begin{equation}
\begin{pmatrix} Z_1 \\ Z_2  \end{pmatrix} = \begin{pmatrix} \cos\phi & - \sin\phi \\
\sin\phi & \cos\phi  \end{pmatrix} \begin{pmatrix} Z_- \\ Z_+  \end{pmatrix} \label{z1z2mixing}
\end{equation}

\begin{equation*}
 \cos\phi \, \sin\phi=\frac{\sqrt{3}}{4} \,\frac{M_2^2}{\sqrt{ M_2^4 + M_3^2 (M_3^2 -  M_2^2) } } 
 \end{equation*}

\vspace{4mm}

\noindent \emph{ Notice that in the limit $y =\frac{M_3}{M_2} \gg 1$, $\sin\phi \rightarrow 0$, $\cos\phi \rightarrow 1$, and we get an approximate $SU(2)$ global symmetry for the $Z_1, Y_1^\pm$ almost degenerated gauge boson masses of order $M_2$. Thus, the hierarchy of scales in the SSB yields an approximate $SU(2)$ global symmetry in the spectrum of $SU(3)$ gauge boson masses. Actually this approximate $SU(2)$ symmetry may play the role of a custodial symmetry to suppress properly the tree level
$\Delta F=2$  "Flavour Changing Neutral Currents" (FCNC) processes mediated by the lower scale of horizontal gauge bosons with masses of few TeV's}

\section{Electroweak symmetry breaking}

Recently ATLAS \cite{ATLAS} and CMS \cite{CMS} at the Large Hadron Collider announced
the discovery of a Higgs-like particle, whose properties, couplings to fermions
and gauge bosons will determine whether it is the SM Higgs or a member of an extended
Higgs sector associated to a BSM theory.  The Electroweak Symmetry Breaking (EWSB) in the
$SU(3)$ family symmetry model involves the introduction of two triplets of $SU(2)_L$
Higgs doublets, namely;

\begin{equation*}
\Phi^u=(3,1,2,-1)=\begin{pmatrix}
\begin{pmatrix} \phi^o\\ \phi^- \end{pmatrix}_1^u \\\\ \begin{pmatrix} \phi^o\\ \phi^- \end{pmatrix}_2^u \\\\
\begin{pmatrix} \phi^o\\ \phi^- \end{pmatrix}_3^u \end{pmatrix} \qquad , \qquad
\Phi^d=(3,1,2,+1)=\begin{pmatrix}
\begin{pmatrix} \phi^+\\ \phi^o \end{pmatrix}_1^d \\\\ \begin{pmatrix} \phi^+\\ \phi^o \end{pmatrix}_2^d \\\\
\begin{pmatrix} \phi^+\\ \phi^o \end{pmatrix}_3^d \end{pmatrix} \, ,
\end{equation*}

\noindent with the VEV?s

\begin{equation*}
 \Phi^u \rangle = \begin{pmatrix}  \langle \Phi_1^u \rangle \\ \langle \Phi_2^u \rangle \\ \langle \Phi_3^u \rangle \end{pmatrix}  \quad , \quad
\langle \Phi^d \rangle= \begin{pmatrix} \langle \Phi_1^d \rangle \\ \langle \Phi_2^d \rangle
\\ \langle \Phi_3^d \rangle \end{pmatrix} \;,
\end{equation*}

\noindent where

\begin{equation*}  \Phi_i^u \rangle = \frac{1}{\sqrt[]{2}}
\begin{pmatrix} v_{ui} \\ 0  \end{pmatrix}  \quad , \quad
\langle \Phi_i^d \rangle = \frac{1}{\sqrt[]{2}}
\begin{pmatrix} 0 \\ v_{di}  \end{pmatrix}  \:.
\end{equation*}

\vspace{3mm}
\noindent The contributions from $\langle \Phi^u \rangle$ and $\langle \Phi^d \rangle$ yield
the $W$ and $Z$ gauge boson masses and mixing with the $SU(3)$ gauge bosons

\begin{multline} \frac{g^2 }{4} \,(v_u^2+v_d^2)\,
W^{+} W^{-} + \frac{ (g^2 + {g^\prime}^2) }{8}  \,(v_u^2+v_d^2)\,Z_o^2   \\
                                                                                                               \\
+ \frac{1}{4} \sqrt{g^2 + {g^\prime}^2} \,g_H\,Z_o \,
\left[ \,(v_{1u}^2-v_{2u}^2 -v_{1d}^2+v_{2d}^2)\,Z_1 + (v_{1u}^2+v_{2u}^2 -2v_{3u}^2 -v_{1d}^2-v_{2d}^2+2v_{3d}^2)\,\frac{Z_2}{\sqrt{3}} \right. \\
                                              \\
\left. + 2\,(v_{1u} v_{2u}-v_{1d} v_{2d})\,\frac{Y_1^+ + Y_1^-}{\sqrt{2}}  + 2\,(v_{1u} v_{3u}-v_{1d} v_{3d})\,\frac{Y_2^+ + Y_2^-}{\sqrt{2}}   + 2\,(v_{2u} v_{3u}-v_{2d} v_{3d})\,\frac{Y_3^+ + Y_3^-}{\sqrt{2}}  \right] \\
                                                                                                    \\
+ \frac{g_H^2}{4} \, \left\{\, \frac{1}{2} \,(v_{1u}^2+v_{2u}^2+v_{1d}^2+v_{2d}^2)\, Z_1^2  +
\frac{1}{2} \,(v_{1u}^2+v_{2u}^2+4 v_{3u}^2+v_{1d}^2+v_{2d}^2+4 v_{3d}^2)\, \frac{Z_2^2}{3} \right. \\
                                                                              \\
+ (v_{1u}^2+v_{2u}^2+v_{1d}^2+v_{2d}^2)\, Y_1^+ Y_1^- + (v_{1u}^2+v_{3u}^2+v_{1d}^2+v_{3d}^2)\, Y_2^+ Y_2^- +(v_{2u}^2+v_{3u}^2+v_{2d}^2+v_{3d}^2) \,Y_3^+ Y_3^-    \\
                                                                 \\
 + (v_{1u}^2-v_{2u}^2 + v_{1d}^2-v_{2d}^2)\,Z_1 \, \frac{Z_2}{\sqrt{3}}                                
+ (v_{2u} v_{3u}+v_{2d} v_{3d})\,(Y_1^+ Y_2^- + Y_1^- Y_2^+)   \\ \\
  + (v_{1u} v_{2u}+v_{1d} v_{2d})\,(Y_2^+ Y_3^- + Y_2^- Y_3^+) +(v_{1u} v_{3u}+v_{1d} v_{3d})\,(Y_1^+ Y_3^+ + Y_1^- Y_3^-)  \\
                                                                                 \\
\left. + 2\,(v_{1u} v_{2u}+v_{1d} v_{2d})\, \frac{Z_2}{\sqrt{3}}\, \frac{Y_1^+ + Y_1^-}{\sqrt{2}} +
(v_{1u} v_{3u}+v_{1d} v_{3d})\, (Z_1 - \frac{Z_2}{\sqrt{3}} )\, \frac{Y_2^+ + Y_2^-}{\sqrt{2}}  \right.  \\  \left.
- (v_{2u} v_{3u}+v_{2d} v_{3d})\, (Z_1 + \frac{Z_2}{\sqrt{3}} )\, \frac{Y_3^+ + Y_3^-}{\sqrt{2}} \right\}  \label{ewyimixcont} \end{multline}

\vspace{5mm}
\noindent $v_u^2=v_{1u}^2+v_{2u}^2+v_{3u}^2$ , $v_d^2= v_{1d}^2+v_{2d}^2+v_{3d}^2$.  Hence,
if we define as usual $M_W=\frac{1}{2} g v$, we may write $ v=\sqrt{v_u^2+v_d^2 } \thickapprox
246$ GeV.

\begin{equation}
Y_j^1=\frac{Y_j^+ + Y_j^-}{\sqrt{2}}  \quad , \quad  Y_j^\pm=\frac{Y_j^1 \mp i Y_j^2}{\sqrt{2}} 
\end{equation}

\vspace{4mm}
\noindent \emph{ The mixing of $Z_o$ neutral gauge boson with the $SU(3)$ gauge bosons modify the couplings of the standard model Z boson with the ordinary quarks and leptons}

\section{ Fermion masses}

\subsection{Dirac See-saw mechanisms}

Now we describe briefly the procedure to get the masses for
fermions. The analysis is presented explicitly for the charged
lepton sector, with a completely analogous procedure for the $u$
and $d$ quarks and Dirac neutrinos. With the fields of particles introduced in
the model, we may write the gauge invariant Yukawa couplings, as

\begin{equation}
h\:\bar{\psi}_l^o \:\Phi^d \:E_R^o \;+\;
h_2 \:\bar{\psi}_e^o \:\eta_2 \:E_L^o \;+\; h_3 \:\bar{\psi}_e^o
\:\eta_3 \:E_L^o \;+\; M \:\bar{E}_L^o \:E_R^o \;+
h.c \label{DiracYC} \end{equation}

\noindent where $M$ is a free mass parameter  because its mass
term is gauge invariant and $h$, $h_2$ and $h_3$ are
Yukawa coupling constants. When the involved scalar fields acquire
VEV's we get, in the gauge basis ${\psi^{o}_{L,R}}^T = ( e^{o} ,
\mu^{o} , \tau^{o}, E^o )_{L,R}$, the mass terms $\bar{\psi}^{o}_L
{\cal{M}}^o \psi^{o}_R + h.c $, where

\begin{equation}
{\cal M}^o = \begin{pmatrix} 0 & 0 & 0 & h \:v_1\\ 0 & 0 & 0 & h \:v_2\\
0 & 0 & 0 & h \:v_3\\
0 & h_2 \Lambda_2  & h_3 \Lambda_3 & M
\end{pmatrix} \equiv \begin{pmatrix} 0 & 0 & 0 & a_1\\ 0 & 0 & 0 & a_2\\
0 & 0 & 0 & a_3\\ 0 & b_2 & b_3 & M
\end{pmatrix} \;. \label{tlmassmatrix} \end{equation}

\noindent Notice that ${\cal{M}}^o$ has the same structure of a
See-saw mass matrix, here for Dirac fermion masses.
So, we call ${\cal{M}}^o$ a {\bf "Dirac See-saw"} mass matrix.
${\cal{M}}^o$ is diagonalized by applying a biunitary
transformation $\psi^{o}_{L,R} = V^{o}_{L,R} \;\chi_{L,R}$. The
orthogonal matrices $V^{o}_L$ and $V^{o}_R$ are obtained
explicitly in Appendix A. From $V_L^o$ and $V_R^o$, and using
the relationships defined there, one computes

\begin{eqnarray}
{V^{o}_L}^T {\cal{M}}^{o} \;V^{o}_R =Diag(0,0,-
\lambda_3,\lambda_4)   \label{tleigenvalues}\\
                                  \nonumber   \\
{V^{o}_L}^T {\cal{M}}^{o} {{\cal{M}}^{o}}^T \;V^{o}_L = {V^{o}_R}^T
{{\cal{M}}^{o}}^T {\cal{M}}^{o} \;V^{o}_R =
Diag(0,0,\lambda_3^2,\lambda_4^2)  \:.\label{tlLReigenvalues}\end{eqnarray}

\noindent where $\lambda_3^2$ and $\lambda_4^2$ are the nonzero
eigenvalues defined in Eqs.(\ref{nonzerotleigenvalues}-\ref{paramtleigenvalues}),
$\lambda_4$ being the fourth heavy fermion mass, and $\lambda_3$ of
the order of the top, bottom and tau mass for u, d and e fermions, respectively.
We see from Eqs.(\ref{tleigenvalues},\ref{tlLReigenvalues}) that at tree level the
See-saw mechanism yields two massless eigenvalues associated to the light fermions.

\section{One loop contribution to fermion masses}

Subsequently, the masses for the light fermions arise through one
loop radiative corrections. After the breakdown of the electroweak
symmetry we can construct the generic one loop mass diagram of
Fig. 1. Internal fermion line in this diagram represent the Dirac see-saw
mechanism implemented by the couplings in Eq.(\ref{DiracYC}). The vertices
read from the $SU(3)$ flavor symmetry interaction Lagrangian

\begin{multline} i {\cal{L}}_{int} = \frac{g_{H}}{2}
\left( \bar{e^{o}}
\gamma_{\mu} e^{o}- \bar{\mu^{o}} \gamma_{\mu} \mu^{o} \right) Z_1^\mu
+  \frac{g_{H}}{2 \sqrt{3}} \left( \bar{e^{o}} \gamma_{\mu} e^{o}+ \bar{\mu^{o}}
\gamma_{\mu} \mu^{o} - 2 \bar{\tau^{o}}
\gamma_{\mu} \tau^{o}  \right) Z_2^\mu                \\
+ \frac{g_{H}}{\sqrt{2}} \left( \bar{e^{o}} \gamma_{\mu} \mu^{o} Y_1^{+} +
\bar{e^{o}} \gamma_{\mu} \tau^{o} Y_2^{+} + \bar{\mu^{o}} \gamma_{\mu} \tau^{o} Y_3^{+} + h.c.
\right) \:,\label{SU3lagrangian} \end{multline}

\begin{figure}[!]
\centering\includegraphics[width=.7\textwidth]{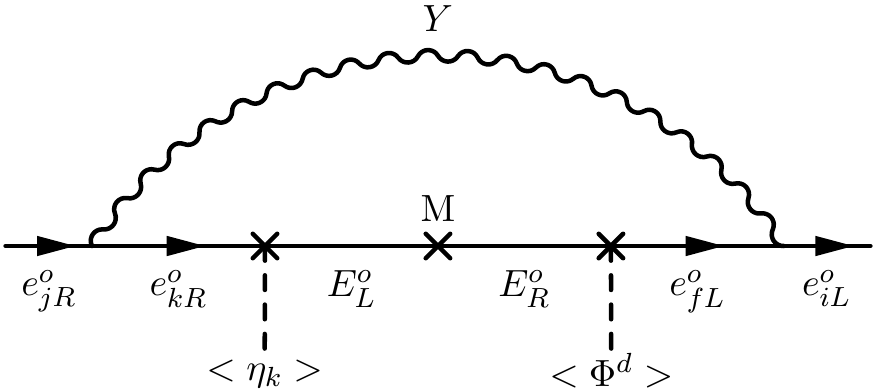}
\caption{ Generic one loop diagram contribution to the mass term
$m_{ij} \:{\bar{e}}_{iL}^o e_{jR}^o$}
\end{figure}

\noindent where $g_H$ is the $SU(3)$ coupling constant, $Z_1$, $Z_2$
and $Y_i^j\;,i=1,2,3\;,j=1,2,$ are the eight gauge bosons. The
crosses in the internal fermion line mean tree level mixing, and
the mass $M$ generated by the Yukawa couplings in Eq.(\ref{DiracYC})
after the scalar fields get VEV's. The one loop diagram of Fig. 1
gives the generic contribution to the mass term $m_{ij}
\:{\bar{e}}_{iL}^o e_{jR}^o$

\begin{equation} c_Y \frac{\alpha_H}{\pi} \sum_{k=3,4} m_k^o
\:(V_L^o)_{ik}(V_R^o)_{jk} f(M_Y, m_k^o) \qquad , \qquad \alpha_H
\equiv \frac{g_H^2}{4 \pi} \end{equation}

\noindent  where $M_Y$ is the gauge boson mass, $c_Y$ is a factor
coupling constant, Eq.(\ref{SU3lagrangian}), $m_3^o=-\sqrt{\lambda_3^2}$ and
$m_4^o=\lambda_4$ are the See-saw mass eigenvalues,
Eq.(\ref{tleigenvalues}), and $f(x,y)=\frac{x^2}{x^2-y^2}
\ln{\frac{x^2}{y^2}}$. Using the results of Appendix A, we
compute

\begin{equation} \sum_{k=3,4} m_k^o \:(V_L^o)_{ik}(V_R^o)_{jk} f(M_Y,
m_k^o)= \frac{a_i \:b_j \:M}{\lambda_4^2 - \lambda_3^2}\:F(M_Y) \:,
\end{equation}

\noindent $i=1,2,3$ , $j=2,3$, and $F(M_Y)\equiv
\frac{M_Y^2}{M_Y^2 - \lambda_4^2} \ln{\frac{M_Y^2}{\lambda_4^2}} -
\frac{M_Y^2}{M_Y^2 - \lambda_3^2} \ln{\frac{M_Y^2}{\lambda_3^2}}$. Adding up all
the one loop $SU(3)$ gauge boson contributions, we get the mass terms
$\bar{\psi^{o}_L} {\cal{M}}_1^o  \:\psi^{o}_R + h.c.$,

\vspace{1mm}

\begin{equation}
{\cal{M}}_1^o = \left( \begin{array}{ccrc} D_{11} & D_{12} & D_{13}  & 0\\
0 & D_{22} & D_{23} & 0\\ 0 & D_{32} & D_{33} & 0\\
0 & 0 & 0 & 0
\end{array} \right) \:\frac{\alpha_H}{\pi}\; ,
\end{equation}

\vspace{1mm}

\begin{eqnarray*}
D_{11}&=&\frac{1}{2} ( \mu_{22} F_1+\mu_{33} F_2 ) \\
D_{12}&=&\mu_{12} (- \frac{F_{Z_1}}{4}+\frac{F_{Z_2}}{12}) \\
D_{13}&=&- \mu_{13} ( \frac{F_{Z_2}}{6}+ F_m ) \\
D_{22}&=&\mu_{22} (\frac{F_{Z_1}}{4}+\frac{F_{Z_2}}{12} - F_m )+\frac{1}{2} \mu_{33} F_3  \\
D_{23}&=&- \mu_{23} ( \frac{F_{Z_2}}{6} - F_m ) \\
D_{32}&=&- \mu_{32} ( \frac{F_{Z_2}}{6} - F_m ) \\
D_{33}&=& \mu_{33} \frac{F_{Z_2}}{3}+\frac{1}{2} \mu_{22} F_3 \:, 
\end{eqnarray*}

\vspace{3mm}

\begin{equation*}
F_1 \equiv F(M_{Y_1}) \quad,\quad F_2 \equiv F(M_{Y_2}) \quad,\quad F_3 \equiv F(M_{Y_3}) \end{equation*}

\begin{equation*}
M_{Y_1}^2=M_2^2 \quad,\quad M_{Y_2}^2=M_3^2 \quad,\quad M_{Y_3}^2=M_2^2+M_3^2 \,
\end{equation*}

\begin{equation*}
F_m=\frac{\cos\phi \sin\phi}{2 \sqrt{3}}\, [\, F(M_-)-F(M_+)\,]
\end{equation*}

\noindent with $M_2, M_3, M_-$ and $M_+$ the  boson masses defined in Eqs.(\ref{M23}-\ref{Mp}).

\vspace{4mm}
\noindent Due to the $Z_1 - Z_2$ mixing, we diagonalize the propagators involving  $Z_1$ and $Z_2$  gauge bosons according to Eq.(\ref{z1z2mixing})

\begin{equation*}
Z_1 = \cos\phi \;Z_- - \sin\phi \;Z_+  \quad , \quad Z_2 = \sin\phi \;Z_- + \cos\phi \;Z_+
\end{equation*}

\begin{eqnarray*}
\langle Z_1 Z_1 \rangle  &=&  \cos^2\phi\; \langle Z_- Z_- \rangle  +  \sin^2\phi\; \langle Z_+ Z_+ \rangle \\\\
\langle Z_2 Z_2 \rangle  &=&  \sin^2\phi\; \langle Z_- Z_- \rangle  +  \cos^2\phi\; \langle Z_+ Z_+ \rangle \\\\
\langle Z_1 Z_2 \rangle  &=&  \sin\phi \, \cos\phi \;( \langle Z_- Z_- \rangle - \langle Z_+ Z_+ \rangle )
\end{eqnarray*}

\noindent So, in the one loop diagram contributions:

\begin{equation*}
F_{Z_1}=\cos^2\phi \,F(M_-) + \sin^2\phi \,F(M_+) \qquad , \qquad  F_{Z_2}=\sin^2\phi \,F(M_-) + \cos^2\phi \,F(M_+) \, ,
\end{equation*}

\begin{equation} \mu_{ij}=\frac{a_i \:b_j \:M}{\lambda_4^2 - \lambda_3^2} = \frac{a_i
\:b_j}{a \:b} \:\lambda_3\:c_{\alpha} \:c_{\beta} \:,\end{equation}

\noindent and $c_{\alpha} \equiv \cos\alpha \:,\;c_{\beta} \equiv \cos\beta \:,\;
s_{\alpha} \equiv \sin\alpha \:,\;s_{\beta} \equiv \sin\beta$, as defined in the
Appendix, Eq.(\ref{Seesawmixing}). Therefore, up to one loop
corrections we obtain the fermion masses

\begin{equation} \bar{\psi}^{o}_L {\cal{M}}^{o} \:\psi^{o}_R + \bar{\psi^{o}_L}
{\cal{M}}_1^o \:\psi^{o}_R = \bar{\chi_L} \:{\cal{M}}
\:\chi_R \:,\end{equation}

\vspace{1mm} 

\noindent with ${\cal{M}} \equiv  \left[ Diag(0,0,-\lambda_3,\lambda_4)+ {V_L^o}^T {\cal{M}}_1^o\:V_R^o \right]$. Using $V_L^o$, $V_R^o$ from Eqs.(\ref{VoL}-\ref{VoR}) we get the mass matrix

\begin{equation} {\cal{M}}= \begin{pmatrix} 
m_{11}&m_{12}&c_\beta \:m_{13}&s_\beta \:m_{13} \\
                                                             \\
m_{21}& m_{22} & c_\beta \:m_{23} & s_\beta \:m_{23}\\
                                                             \\
c_\alpha \:m_{31}& c_\alpha \:m_{32} & (-\lambda_3+c_\alpha c_\beta
\:m_{33}) & c_\alpha s_\beta \:m_{33} \\
                                           \\
s_\alpha \:m_{31}& s_\alpha \:m_{32} & s_\alpha c_\beta \:m_{33} &
(\lambda_4+s_\alpha s_\beta \:m_{33})
\end{pmatrix}  \;,\label{massVI}
\end{equation}

\vspace{4mm} 
\noindent where 

\begin{eqnarray}
m_{11}=\frac{1}{2} \frac{a_2}{a^\prime}  \Pi_1 \quad ,& \quad m_{12}= - \frac{1}{2} \frac{a_1 b_3}{a^\prime b} (  \Pi_2 -6 \mu_{22} F_m )  \\
                    \nonumber \\
m_{21}= \frac{1}{2} \frac{a_1 a_3}{a^\prime a}\Pi_1  \quad ,& \quad
m_{31}=\frac{1}{2} \frac{a_1}{a}  \Pi_1 \end{eqnarray}

\begin{equation} m_{13}=- \frac{1}{2} \frac{a_1 b_2}{a^\prime b} [\Pi_2 +2(2\frac{b_3^2}{b_2^2}-1)
\mu_{22}F_m ]   \end{equation}

\begin{equation}
m_{22}=\frac{1}{2} \frac{a_3 b_3}{a \, b} \left[\frac{a_2}{a^\prime} ( \Pi_2 -6 \mu_{22} F_m )+
\frac{a^\prime b_2}{a_3 b_3} (  \Pi_3 + \Delta )        \right]
\end{equation}

{\footnotesize
\begin{equation} 
m_{23}=\frac{1}{2} \frac{a_3 b_3}{a \, b}  \left[\frac{a_2 b_2}{a^\prime b_3} ( \Pi_2 +2(2\frac{b_3^2}{b_2^2}-1 ) \mu_{22} F_m ) - \frac{a^\prime}{a_3} (  \Pi_3 -\frac{b_2^2}{b_3^2} \Delta +2\frac{b^2}{b_3^2}\mu_{33} F_m    )        \right] 
\end{equation}
}

\begin{equation} m_{32}=\frac{1}{2} \frac{a_3 b_3}{a \, b}  \left[\frac{a_2}{a_3} ( \Pi_2 -6 \mu_{22} F_m)-\frac{b_2}{b_3} (  \Pi_3 -\frac{{a^\prime}^2 }{a_3^2} \Delta -2\frac{a^2}{a_3^2}\mu_{33} F_m ) \right] \end{equation}

\begin{equation}
m_{33}=\frac{1}{2} \frac{a_3 b_3}{a \, b} \left[\frac{a_2 b_2}{a_3 b_3} ( \Pi_2 - 2 \mu_{22} F_m ) + \Pi_3+ \frac{ {a^\prime}^2 b_2^2}{a_3^2 b_3^2} \Delta - \frac{1}{3} \frac{a^2 b^2}{a_3^2 b_3^2}\mu_{33} F_{Z_2}  
+ 2 ( \frac{b_2^2}{b_3^2} + 2\frac{a_2^2}{a_3^2}-\frac{{a^\prime}^2}{a_3^2} )\mu_{33} F_m          \right]
\end{equation}

\vspace{3mm}
\begin{eqnarray}
\Pi_1 = \mu_{22} F_1 +  \mu_{33} F_2 \quad ,& \quad \Pi_2 = \mu_{22} F_{Z_1} +  \mu_{33} F_3 \nonumber \\
                                                                             \nonumber \\
\Pi_3 = \mu_{22} F_3 +  \mu_{33} F_{Z_2} \quad ,& \quad \Delta = \frac{1}{2}\mu_{33}(F_{Z_2} - F_{Z_1} )
\end{eqnarray}

\vspace{4mm}
\noindent \emph{Notice that the $m_{ij}$ mass terms depend just on the 
$\frac{a_i}{a_j}$ and $\frac{b_i}{b_j}$ ratios of the tree level parameters.}

\vspace{5mm}

\begin{equation} a^\prime=\sqrt{a_1^2+a_2^2}\;\; , \;\;a=\sqrt{{a^\prime}^2+a_3^2} \;\; ,
\;\; b=\sqrt{{b_2^2+b_3^2}} \;, 
\end{equation}

\vspace{4mm}

\noindent The diagonalization of ${\cal{M}}$,
Eq.(\ref{massVI})  gives the physical masses for  u,
d, and e charged fermions. Using a new biunitary transformation
$\chi_{L,R}=V_{L,R}^{(1)} \;\Psi_{L,R}$;
\;$\bar{\chi}_L \;{\cal{M}} \;\chi_R= \bar{\Psi}_L \:{V_L^{(1)}}^T
{\cal{M}} \; V_R^{(1)} \:\Psi_R $, with ${\Psi_{L,R}}^T = ( f_1 ,
f_2 , f_3 , F )_{L,R}$ the mass eigenfields, that is

\begin{equation}
{V^{(1)}_L}^T {\cal{M}} \:{\cal M}^T \;V^{(1)}_L =
{V^{(1)}_R}^T {\cal M}^T \:{\cal{M}} \;V^{(1)}_R =
Diag(m_1^2,m_2^2,m_3^2,M_F^2) \:,\end{equation}

\noindent $m_1^2=m_e^2$, $m_2^2=m_\mu^2$, $m_3^2=m_\tau^2$ and
$M_F^2=M_E^2$ for charged leptons.

\vspace{5mm}
\subsection{Quark $( V_{CKM} )_{4\times 4}$ mixing matrix }

Within this $SU(3)$ family symmetry model, the transformations from
massless to physical mass fermion eigenfields for quarks and charged leptons are

\begin{equation*} \psi_L^o = V_L^{o} \:V^{(1)}_L \:\Psi_L \qquad \mbox{and}
\qquad \psi_R^o = V_R^{o} \:V^{(1)}_R \:\Psi_R \,.\end{equation*}

\noindent Recall  that vector like quarks, Eq.(\ref{vectorquarks}), are $SU(2)_L$
weak singlets, and hence they do not couple to the $W$ boson in the
interaction basis. In this way, the interaction of  L-handed up and down quarks; ${f_{uL}^o}^T=(u^o,c^o,t^o)_L$ and ${f_{dL}^o}^T=(d^o,s^o,b^o)_L$, to the $W$ charged gauge boson may be written as

\begin{equation} \frac{g}{\sqrt{2}} \,\bar{f^o}_{u L} \gamma_\mu f_{d L}^o
{W^+}^\mu = \frac{g}{\sqrt{2}} \,\bar{\Psi}_{u L}\;
[(V_{u L}^o\,V_{u L}^{(1)})_{3\times 4}]^T \;(V_{d L}^o\,V_{d L}^{(1)})_{3\times 4}\;
\gamma_\mu \Psi_{d L} \;{W^+}^\mu \:,\end{equation}

\noindent where $g$ is the $SU(2)_L$ gauge coupling. Therefore, the non-unitary $V_{CKM}$ of dimension $4\times4$ is identified as

\begin{equation} (V_{CKM})_{4\times 4} = [(V_{u L}^o\,V_{u L}^{(1)})_{3\times 4}]^T \;(V_{d L}^o\,V_{d L}^{(1)})_{3\times 4}
\end{equation}

\pagebreak
\section{Numerical results} \label{numerical}

\emph{To illustrate the spectrum of masses and mixing, let us consider the following
fit of space parameters at the $M_Z$ scale \cite{xingzhang}}

\vspace{3mm}
\noindent Taking the input values

\begin{equation*} M_2 = 2\,\text{TeV} \quad , \quad M_3 = 2000\,\text{TeV} \quad , \quad \frac{\alpha_H}{\pi}=0.2 \end{equation*}

\noindent for the $M_2$, $M_3$ horizontal boson masses, Eq.(\ref{M23}), and the $SU(3)$  coupling
constant, respectively, and the ratio of the electroweak VEV's: $v_{iu}$ from $\Phi^u\;$  ($v_{id}$ from $\Phi^d$)

\begin{equation*} v_{1u}=0  \quad , \quad \frac{ v_{2u}}{ v_{3u}} = 0.1  \quad , \quad
\frac{ v_{1d}}{ v_{2d}} = 0.23257  \quad , \quad  \frac{ v_{2d}}{ v_{3d}}=0.08373 \: ,
\end{equation*}

\noindent we obtain the following mass and mixing matrices, and mass eigenvalues: 

\subsection{Quark masses and mixing}

\vspace{5mm}
\noindent {\bf u-quarks:}

\vspace{3mm}
\noindent Tree level see-saw mass matrix:

\begin{equation} {\cal M}_u^o=
\left(
\begin{array}{cccc}
 0 & 0 & 0 & 0. \\
 0 & 0 & 0 & 29834. \\
 0 & 0 & 0 & 298340. \\
 0 & 1.49495\times 10^7 & -730572. & 1.58511\times 10^7 \\
\end{array}
\right)
\,\text{MeV} \,,\end{equation}

\noindent the mass matrix up to one loop corrections:

\begin{equation} {\cal M}_u=
\left(
\begin{array}{cccc}
 1.38 & 0. & 0. & 0. \\
 0. & -532.587 & -2587.14 & -2442.42 \\
 0. & 7064.64 & -172017. & 31927.1 \\
 0. & 70.6499 & 338.204 & 2.18023\times 10^7 \\
\end{array}
\right)\,\text{MeV} \, ,
\end{equation}

\noindent and the u-quark masses

\begin{equation}
(\,m_u \;,\; m_c \;,\; m_t \;,\; M_U\,)=
(\,1.38\;,\; 638.22 \;,\;172181\;,\;2.18023\times 10^7\,)\,\text{MeV}
\end{equation}

\vspace{7mm}
\noindent {\bf d-quarks:}

\vspace{3mm}
\begin{equation} {\cal M}_d^o=
\left(
\begin{array}{cccc}
 0 & 0 & 0 & 13375.7 \\
 0 & 0 & 0 & 57510.3 \\
 0 & 0 & 0 & 686796. \\
 0 & 723708. & -37338.1 & 6.89219\times 10^7 \\
\end{array}
\right)\;\text{MeV} 
\end{equation}

\vspace{3mm}
\begin{equation} {\cal M}_d=
\left(
\begin{array}{cccc}
 2.82461 & 0.0338487 & -0.656039 & -0.00689715 \\
 0.65453 & -25.1814 & -217.369 & -2.28527 \\
 0.0562685 & 423.166 & -2820.62 & 46.5371 \\
 0.000562713 & 4.23187 & 44.2671 & 6.89291\times 10^7 \\
\end{array}
\right)
\;\text{MeV} 
\end{equation}

\vspace{3mm}
\begin{equation}
(\,m_d \;,\; m_s \;,\; m_b \;,\; M_D\,)=
(\, 2.82368  \;,\; 57.0005\;,\; 2860   \;,\; 6.89291\times 10^7 \,)\;\text{MeV}
\end{equation}

\noindent and the quark mixing

\begin{equation} V_{CKM}=
\left(
\begin{array}{cccc}
 0.97362 & 0.225277 & -0.0362485 & 0.000194044 \\
 -0.226684 & 0.973105 & -0.040988 & -0.000310055 \\
 0.0260403 & 0.0481125 & 0.998387 & -0.00999333 \\
 -0.000234396 &- 0.000826552 & -0.011432 & 0.000114632 \\
\end{array}\right)
\label{vckm} \end{equation}

\vspace{5mm}
\subsection{Charged leptons:}

\vspace{3mm}
\begin{equation} {\cal M}_e^o=
\left(
\begin{array}{cccc}
 0 & 0 & 0 & 37956.9 \\
 0 & 0 & 0 & 189784. \\
 0 & 0 & 0 & 1.93543\times 10^6 \\
 0 & 548257. & -30307.4 & 1.94497\times 10^8 \\
\end{array}\right)\;\text{MeV}
 \end{equation}

\vspace{3mm}
\begin{equation} {\cal M}_e=
\left(
\begin{array}{cccc}
 -0.486368 & -0.00536888 & 0.0971221 & 0.000274163 \\
 -0.0967909 & -34.7536 & -250.305 & -0.706579 \\
 -0.0096786 & 485.768 & -1661.27 & 10.8107 \\
 -0.0000967909 & 4.85792 & 38.2989 & 1.94507\times 10^8 \\
\end{array}
\right)\;\text{MeV} 
\end{equation}

\noindent fit the charged lepton masses:

\begin{equation*}
( m_e \,,\, m_\mu \,,\, m_\tau \,,\, M_E ) = ( 0.486095 \,,\,102.7\,,\,1746.17\,,\, 3.15956\times 10^8\, )\,\text{MeV}
\end{equation*}

\noindent and the charged lepton mixing

\begin{equation} V_{e \,L}^o\, V_{e \,L}^{(1)}=
\left(
\begin{array}{cccc}
 0.973942 &   0.221206 & 0.050052 & 0.000194 \\
 -0.226798 & 0.949931 & 0.214927 & 0.0008342 \\
-2.90427\times 10^{-6} & -0.220675 & 0.975296 & 0.009963 \\
 2.62189\times 10^{-7} & 0.0013632 & -0.009906& 0.99995 \\
\end{array}
\right)
\label{emix} \end{equation}

\pagebreak
\section{Conclusions}

We reported recent numerical analysis on charged fermion masses and mixing within a BSM with a local $SU(3)$ family symmetry, which combines tree level "Dirac See-saw" mechanisms and radiative corrections to implement a successful hierarchical mass generation mechanism for quarks and charged leptons.

\noindent In section \ref{numerical} we show a parameter space region where this scenario account
for the known hierarchical spectrum of ordinary quarks and charged lepton masses, and the quark mixing 
in a non-unitary $(V_{CKM})_{4\times 4}$ within allowed values\footnote{except $(V_{CKM})_{13}$ and $(V_{CKM})_{31}$} reported in PDG 2014 \cite{PDG2014}.

\vspace{4mm}
\emph{Let me point out here that the solutions for fermion masses and mixing reported
in section \ref{numerical} suggest that the dominant contribution to EWSB
comes from the weak doublets which couple to the third family.}

\vspace{3mm}
\emph{It is also worth to comment that fermion content, scalar fields, and their transformation 
under the gauge group, Eq. \eqref{gaugegroup}, all together, forbid tree level Yukawa couplings between
ordinary standard model fermions. Consequently, the flavon scalar fields introduced to break
the symmetries: $\Phi^u$, $\Phi^d$, $\eta_2$ and $\eta_3$, couple only
ordinary fermions to their corresponding vector like fermion at tree level. Thus, FCNC
scalar couplings to ordinary fermions are suppressed by light-heavy mixing angles,
which as is shown in the quark mixing $(V_{CKM})_{4 \times 4}$, Eq.(\ref{vckm}), and the charged lepton mixing, Eq. \eqref{emix}, may be small enough to suppress properly the FCNC mediated by the scalar fields within this scenario.}

\vspace{5mm}
\section*{Acknowledgements}

It is a pleasure to thank the organizers N.S. Mankoc-Borstnik, H.B. Nielsen, M. Y. Khlopov,
and participants for the stimulating Workshop at Bled, Slovenia. This work was
partially supported by the "Instituto Polit\'ecnico Nacional",
(Grants from EDI and COFAA) and "Sistema Nacional de
Investigadores" (SNI) in Mexico.

\appendix

\section{Diagonalization of the generic Dirac See-saw mass matrix}

\begin{equation} {\cal M}^o=
\begin{pmatrix} 0 & 0 & 0 & a_1\\ 0 & 0 & 0 & a_2\\ 0 & 0 & 0 &
a_3\\ 0 & b_2 & b_3 & c \end{pmatrix} \end{equation}

\vspace{1mm} \noindent Using the biunitary transformations
$\psi^{o}_L = V_L^o \:\chi_L$ and  $\psi^{o}_R = V_R^o
\:\chi_R $ to diagonalize ${\cal{M}}^o$, the orthogonal matrices
$V^{o}_L$ and $V^{o}_R$ may be written explicitly as 

\begin{equation} V^{o}_L = \begin{pmatrix} \frac{a_2}{a^\prime}& \frac{a_1 a_3}{a^\prime
a} & \frac{a_1}{a} \cos\alpha &
\frac{a_1}{a} \sin\alpha\\
                        \\
- \frac{a_1}{a^\prime}  & \frac{a_2 a_3}{a^\prime a}  &
\frac{a_2}{a} \cos\alpha &
\frac{a_2}{a} \sin\alpha\\
                        \\
0 & - \frac{a^\prime}{a}   & \frac{a_3}{a} \cos{\alpha}
& \frac{a_3}{a} \sin{\alpha}\\
                            \\
0 & 0 & -\sin{\alpha} & \cos{\alpha}
\end{pmatrix}  \label{VoL}
\end{equation}

\begin{equation}
V^{o}_R = \begin{pmatrix}
1 & 0 & 0 & 0 \\
                         \\
0 & \frac{b_3}{b} & \frac{b_2}{b} \cos{\beta} & \frac{b_2}{b} \sin{\beta}\\
                                                     \\
0& - \frac{b_2}{b} & \frac{b_3}{b} \cos{\beta} & \frac{b_3}{b} \sin{\beta}\\
                         \\
0 & 0 & -\sin{\beta} & \cos{\beta}
\end{pmatrix}   \label{VoR}
 \end{equation}

\vspace{3mm}

\noindent where

\begin{equation} \lambda_3^2 = \frac{1}{2} \left( B - \sqrt{B^2 -4D} \right) \quad , \quad  \lambda_4^2 = \frac{1}{2} \left( B + \sqrt{B^2 -4D} \right)
\label{nonzerotleigenvalues}
\end{equation}

\vspace{3mm}
\noindent are the nonzero eigenvalues of
${\cal{M}}^{o} {{\cal{M}}^{o}}^T$ (${{\cal{M}}^{o}}^T
{\cal{M}}^{o}$), and

\begin{eqnarray} B = a^2 + b^2 + c^2 =
\lambda_3^2+\lambda_4^2\quad &, \quad D= a^2
b^2=\lambda_3^2\lambda_4^2 \;,\label{paramtleigenvalues} \end{eqnarray}

\vspace{1mm}

 \begin{equation} \cos{\alpha} =\sqrt{\frac{\lambda_4^2 -
a^2}{\lambda_4^2 - \lambda_3^2}} \quad , \quad \sin{\alpha} =
\sqrt{\frac{a^2 - \lambda_3^2}{\lambda_4^2 - \lambda_3^2}}   \quad , \quad \cos{\beta} =\sqrt{\frac{\lambda_4^2 - b^2}{\lambda_4^2 - \lambda_3^2}} \quad , \quad \sin{\beta} = \sqrt{\frac{b^2 -
\lambda_3^2}{\lambda_4^2 - \lambda_3^2}}    \label{Seesawmixing}      
\end{equation}

\end{document}